\documentclass[12pt,letterpaper]{article}
\usepackage[paper=letterpaper,margin=1in]{geometry}

\usepackage{amsmath,amssymb}

\usepackage{changepage}

\usepackage[utf8x]{inputenc}

\usepackage{textcomp,marvosym}

\usepackage{cite}

\usepackage{nameref,hyperref}

\usepackage[right]{lineno}

\usepackage{microtype}
\DisableLigatures[f]{encoding = *, family = * }

\usepackage[table]{xcolor}

\usepackage{array}

\newcolumntype{+}{!{\vrule width 2pt}}

\newlength\savedwidth

\usepackage[aboveskip=1pt,labelfont=bf,labelsep=period,singlelinecheck=off]{caption}

\makeatletter
\renewcommand{\@biblabel}[1]{\quad#1.}
\makeatother

\usepackage{lastpage,fancyhdr,graphicx}
\usepackage{epstopdf}

\newcommand{\theversion}%
   {{version = 5.41 of lattice.tex 2019 Oct 29}}
\newcommand{\fig}[1]{Fig~\ref{#1}} %
\newcommand{\figmargin}[1]{} %
\newcommand{\eq}[1]{(\ref{#1})} %
\newcommand{\tablereference}[1]{Table \ref{#1}} %
\newcommand{\tablemargin}[1]{} %
\newcommand{\kb}{k_{\text{B}}} %
\newcommand{\ds}{d_{\text{space}}} %
\newcommand{\rfrequency}{R_{\text{frequency}}}
\newcommand{\rsequence}{R_{\text{sequence}}}
\newcommand{\deltaGnaught}{ \Delta G^{\circ} } %
\newcommand{\renergy}{R_\text{energy}}
\newcommand{\Emin}{E_{\text{min}}} %

\begin{document}
\vspace*{0.2in}

\begin{flushleft}
{\Large\textbf{%
Restriction enzymes
use a 24 dimensional coding space
to recognize 6 base long DNA sequences
}
}
\makeatletter{\renewcommand*{\@makefnmark}{}
\footnotetext{\theversion}\makeatother}
\newline
\\
Thomas D. Schneider\textsuperscript{1},
Vishnu Jejjala\textsuperscript{2,3}
\\
\bigskip
\textbf{1}
National Institutes of Health,
National Cancer Institute, Center for Cancer Research,
RNA Biology Laboratory,
Frederick, Maryland,
United States of America,
schneidt@mail.nih.gov
\\
\vspace{6pt}
\textbf{2}
Mandelstam Institute for Theoretical Physics,
School of Physics, NITheP, and CoE-MaSS,
University of the Witwatersrand,
Johannesburg, South Africa,
vishnu@neo.phys.wits.ac.za
\\
\vspace{6pt}
\textbf{3}
David Rittenhouse Laboratory,
University of Pennsylvania,
Philadelphia, Pennsylvania,
United States of America
\\
\bigskip

\end{flushleft}
\section*{Abstract}
Restriction enzymes recognize and
bind to specific sequences on invading
bacteriophage DNA.
Like a key in a lock, these proteins
require many contacts
to specify the correct DNA sequence.
Using information theory
we develop an equation that defines
the number of independent contacts, which is the dimensionality
of the binding.  We show that EcoRI,
which binds to the sequence GAATTC,
functions in 24 dimensions.
Information theory
represents messages as spheres in high dimensional spaces.
Better sphere packing leads to better communications
systems.  The densest known packing of hyperspheres occurs on the Leech lattice
in 24 dimensions.  We suggest that the single protein
EcoRI molecule employs a Leech lattice in its operation.
Optimizing density of sphere packing explains why 6 base restriction enzymes
are so common.

\newpage %

\section{Introduction}
\label{sec:intro}

Restriction
enzymes provide a defense mechanism in procaryotes
against foreign DNA injected by
bacteriophages~\cite{Roberts2005,Pingoud.Wende2014}.
These proteins bind to specific sequences on DNA and cleave the DNA,
rendering it susceptible to attack by
exonucleases~\cite{Simmon.Lederberg1972, Heitman.Model1989}
and preventing viral replication.  The
bacterial genome is protected by modification enzymes that methylate
the same pattern that the restriction enzymes cut.  Though the
sequences bound by restriction enzymes usually consist of only $4$ or
$6$ base pairs,
even a single base change
of the GAATTC EcoRI
binding site decreases EcoRI binding by at least 1000 fold
\cite{Lesser.Jen-Jacobson1990}.
How can
restriction enzymes have such precise recognition?  Why do we find the
majority of restriction enzymes have exactly $4$ or $6$ base pair long
recognition sequences?

We begin with a brief overview of how
concepts from information and coding theory can
be used to answer these questions.
We address this paper to
both biologists and information/coding
theorists.
Therefore some of this material may be familiar to one audience and 
foreign to the other.

In this paragraph, we state our main result.
We model the binding process of restriction enzymes
as a selection between distinct states.
These states can be represented as spheres packed together in a high
dimensional `coding' space, and the equations of information theory along
with empirical data for EcoRI binding allow us to determine the
dimensionality of the space.  Surprisingly, operating at the maximum
biological efficiency,
the $6$-base cutting enzyme EcoRI
works in a $24$ dimensional space, and so is likely to use the best
sphere packing known, the famous $24$ dimensional Leech
lattice~\cite{Leech1964}. 
Using the Leech lattice would allow restriction enzymes to
minimize their errors and so cut precisely.
The $4$-base cutters should work in a $16$ dimensional space,
and in that dimension there are also good sphere packings.
Apparently, restriction enzymes have evolved
over a high dimensional landscape
to take advantage
of the best sphere packings.

To model recognition of EcoRI binding to DNA,
we distinguish two different energy flows in time.
The first is the energy dissipated
during the binding or `operation'
of the molecules
$P$ (power)
\cite{Schneider.ccmm}.
To describe $P$ for EcoRI we consider two states,
and in both states the protein is associated with the DNA.
The dissipation of $P$ proceeds from a
high energy `before' state in which the EcoRI molecule
is somewhere on the DNA but not bound specifically.
Then, after a Brownian motion search
\cite{Weber.Steitz1984a,Piatt.Price2019}
when EcoRI encounters a binding site it may
begin to form specific bonds \cite{Weber.Steitz1984b, McClarin.Rosenberg1986}.
As these bonds form,
energy is dissipated to the surrounding water until
EcoRI has formed all its bonds to the DNA.  We call this
latter low energy state the `after' state.
The energetic difference between these
two states is the
specific binding energy,
$P = -\Delta G^{\circ}_{\text{spec}}$
with units of joules per binding
\cite{Schneider-emmgeo2010}.
The time of this dissipation may vary, but the total energy
dissipated is constant between the two states,
as indicated by the measurability of
$\Delta G^{\circ}_{\text{spec}}$
for the operation \cite{Clore.Davies1982}.
The second important energy flow involved in EcoRI binding
is the thermal noise that passes through the molecule
\emph{during the binding operation}.
This noise, $N$, interferes with bond formation.
These concepts are parallel to the communications
model developed by Shannon in which the power $P$
of the communication signal is absorbed
into and then dissipates from the receiver
while it selects
a particular message
\cite{Shannon1948}.
The receiver must also handle additional energy
caused by thermal noise $N$ added to the power.
The $P/N$ ratio is called the `signal-to-noise' ratio, but
this term is not appropriate for EcoRI since there is
no external signal.  However, in both models
there are the two energies
dissipated in time, $P$ and $N$,
and
there is a selection of specific states.
Of course, a sufficiently strong thermal noise
will eventually dislodge EcoRI
from its specific binding, but this reversal is
not the selection process we are interested in.
Having set up these concepts allows us to apply
powerful theorems from information theory
to the recognition problem
\cite{Shannon1949, Schneider.ccmm, Schneider.edmm, Schneider-emmgeo2010}.

However, the problem of recognition
cannot be explained by thinking about the protein-DNA
contact
as a single 
interaction.
Instead, there are multiple
interactions including
hydrogen, van der Waals and electrostatic
bonds.
To describe this set of interactions takes a series of numbers.
Some of these interactions
could be
independent like the pins in a lock.
As in a lock, it is advantageous for the pins to be
independent because that way the lock can represent more combinations
and is more secure \cite{Macaulay1998}.
Unlike a lock,
the microscopic EcoRI molecule is continuously impacted and
violently jostled by thermal noise ($N$).
Each molecular `pin' has a velocity that is the sum of many small
impacts, so the central limit theorem from statistics tells us
that the velocity will be approximately Gaussian
\cite{Uhlenbeck.Ornstein1930}.
So the moving parts of a molecule--the `pins'-- that help
EcoRI select GAATTC can be
modeled as a set of independent molecular oscillators
moving under the influence of thermal noise
\cite{Schneider.ccmm}.

EcoRI potentially has many pins, each with
a particular velocity.
When one has a set of independent numbers they can be described
as a point in a high dimensional space.
Furthermore,
when two independent Gaussian distributions are combined
at right angles to represent their independence,
the resulting 2-dimensional distribution is circular
\cite{Schneider.shannonbiologist2006}.
With three independent Gaussian distributions
the combined distribution is a sphere and
when there are more than three the distribution is still
spherical,
a hypersphere.
The radius of the sphere is proportional to the square root
of the thermal noise impacting on the molecule
\cite{Schneider.ccmm}.

The higher the dimension of the sphere, the more the distribution
converges to a single radius and the sphere skin or `thickness'
becomes smaller \cite{Schneider.ccmm}.
To see this, consider
the volume of a ball (the region enclosed by a sphere)
embedded in a $D$ dimensional space,
\begin{equation}
V_D(r) =
\frac {\pi^{\frac{D}{2}}}
      {\Gamma \left(\frac{D}{2} + 1 \right)}
    r^{D} ~.
\label{eqn.V}
\end{equation}
For a radius $r$,
let half the volume lie in the shell between $r_* < r$ and $r$, that is
$\frac12 =
\frac{V_D(r_*)}{V_D(r)} =
\frac{r_*^{D}}{r^{D}}$.
Rearranging gives $r_* =
(\frac12)^{1/D} r$ and as $D\to\infty$, $r_*\to r$, so the volume
is densest near the surface.

\begin{figure}[tbhp]%
\vspace{-0.3cm} 
\centering %
\scalebox{1.00}{\includegraphics*{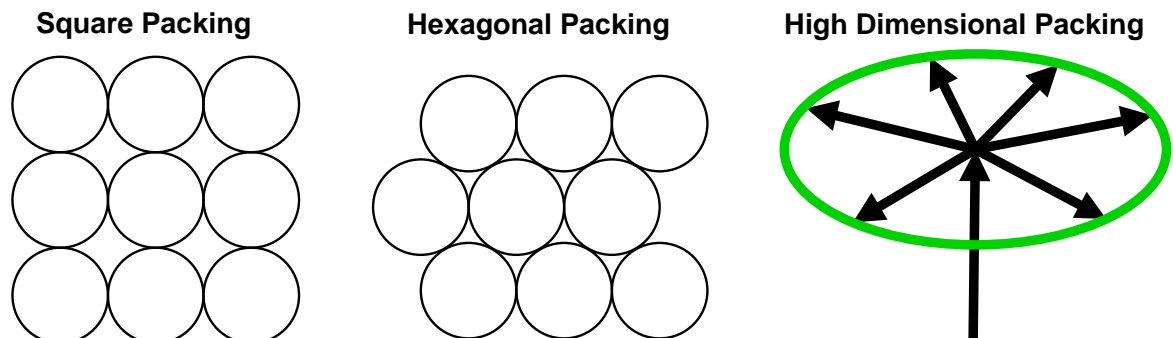}} %
\caption{Sphere packing.
Circles demonstrate square and hexagonal
sphere packing in two dimensions.
The hexagonal packing is $12$\% more dense.
In higher dimensional spaces sphere packing is less intuitive.
When hyperspheres pack together there is an odd property
diagramed on the right side of the figure (which is derived
from Shannon's proof of the channel capacity theorem,
Theorem 2 in his figure 5
\cite{Shannon1949}).
The vertical arrow represents moving
from the center of one hypersphere to the center
of a second hypersphere.
For Shannon,
working with electrical communications,
this voltage is proportional to the square
root of the power dissipation, $\sqrt{P}$.
In a 100 dimensional space, the thermal noise in the second
sphere (green circle) disturbs the signal in all directions,
shown by splayed arrows with lengths proportional to $\sqrt N$.
However, 99 of those dimensions do not perturb
in the direction of the power dissipation.
In his proof, Shannon neglected the 1\% of the noise in
the direction of the power since this represents the error,
and it can be made as small as one may desire by increasing
the dimensionality---in 1000 dimensions the error is only 0.1\%.
So relative to the direction of the power, the received
hypersphere can be treated as a flat surface since all the
other directions (splayed arrows) are at right angles
to the power direction.
If two hyperspheres are to be separated with as low
an error as desired,
then the power to get from one to the next must just
exceed the thermal noise power of the first sphere,
so $\sqrt{P} > \sqrt{N}$
and $P > N$.
}
\vspace{-0.3cm} 
\label{fig.packing}
\end{figure}

So the state of EcoRI bound to GAATTC after dissipation
can be represented as a hypersphere.
Because the pins of EcoRI have an instantaneous position and velocity,
at any one instant EcoRI is at a particular point on the sphere
and moves by Brownian motion across the sphere surface.
EcoRI bound to a different DNA sequence, such as CAATTC,
is on a different hypersphere.
If these two spheres were to intersect, then EcoRI would
be able to bind sites other than GAATTC and this error would be
fatal to the bacterium whose DNA is only protected at GAATTC.
Thus the hyperspheres should not intersect.
When EcoRI binds to DNA,
that provides a finite amount energy that can be dissipated per binding
($P$), so there is a finite set of hyperspheres that can be bound.
During evolution
EcoRI will tend to minimize the binding energy, while the number
of hypersphere states it selects between remains constant
\cite{Schneider-emmgeo2010}
so the hyperspheres
become tightly packed together without intersecting
(\fig{fig.packing}).
\figmargin{fig.packing}
Thus EcoRI can evolve to bind efficiently, using the minimum energy
to select between the maximum number of binding states.
In addition, by using many interactions in a high dimensional space,
the hyperspheres become sharply defined
because the distribution around the sphere radius
(the thickness) becomes smaller
\cite{Schneider.ccmm}.
This allows EcoRI to evolve to reduce
the number of times
it cuts the wrong sequence, giving it a low error rate.

Shannon dealt with a closely related problem regarding
maximizing the information that could
be sent over a phone line for a given power ($P$, joules per second)
\cite{Shannon1948,Shannon1949}.
Messages from a transmitter
can be broken into a series of independent voltages and so the
set of numbers describing a particular message can be represented
as a point in a high dimensional
space which we call the `coding space' since the message
is represented by a code,
the set of voltage values along each dimension.
In addition, thermal noise on the phone wire
causes the received voltage pulses to vary according
to a Gaussian distribution.
So if a message were repeated many times the received
message points would form a sphere in the high dimensional space.
When the receiver gets one of these noise-disturbed points,
it can determine which of the possible transmitted messages is closest
and
thereby
`decode' the message to produce a clear noise-free signal
for the person.
Shannon recognized that the received spheres
should not intersect if the receiver is to avoid
ambiguity in decoding.

The receiver in a communications system selects particular
symbols from all possible symbols that the transmitter might send.
Similarly,
a molecule such as EcoRI selects a particular
state (binding to GAATTC) from an array of possible states
(binding to any arbitrary 6 base long sequence).
This concept applies to many other biological macromolecules.
These `molecular machines'
include proteins that bind DNA, proteins that detect light such
as rhodopsin in the eye
and proteins that cause motion such as myosin moving on actin
in muscle \cite{Schneider.ccmm}.
In every case the molecular machines dissipate energy in order to settle
into one of several possible lower energy states
and they do this despite the presence of violent thermal noise.

This paper answers the question of
why restriction enzymes have such high fidelity
despite being disturbed by thermal noise
by showing
how to calculate the coding space dimensionality of
nucleic-acid recognizing molecular machines.
The measured dimensionalities imply
that restriction enzymes have evolved to exploit coding techniques
only recently developed for modern communication systems.
This in turn suggests that humans should also be able to
build nanometer scale molecules that decode signals.

Background concepts important for understanding these results are
basic information theory~\cite{Pierce1980,SchneiderPrimer}, general
molecular biology~\cite{Watson1987}, and how to measure the
information content of binding sites on DNA or RNA in
bits~\cite{Schneider.Ehrenfeucht1986, Schneider.ev2000}
as in
sequence logos~\cite{Schneider.Stephens1990}.
In addition,
messages in a communications system and the states of molecular
machines can be represented by spheres packing together in a high
dimensional coding space~\cite{Shannon1949, Conway.Sloane1998,
Schneider.ccmm}.
The isothermal efficiency is described in
\cite{Schneider-emmgeo2010, Schneider-brmit2010}.
For reviews, see \cite{Schneider.nano2,Schneider2006,
Schneider-brmit2010}.

The organization of this paper is as follows.
Section \ref{sec:lower}
develops an equation for a lower bound on the dimensionality
and we show how it applies to restriction enzymes in
Section \ref{sec:apply.lower}.
Section \ref{sec:upper} then develops an upper bound,
and
Section \ref{sec:pincers} shows that these bounds converge to
a unique dimension, twice the number of bits.
In Section \ref{sec:dimensionality}
we analyze the dimensionality of over 4000 restriction enzymes
and find that the 4- and 6- base cutting restriction enzymes prefer to
operate in $16$ and $24$ dimensions.  These dimensions contain
dense hypersphere packings.
Section \ref{sec:lattices}
discusses the implications of restriction enzymes using
good sphere packings and examines possible biological reasons
why some restriction enzymes use packings in other dimensions.
Section \ref{sec:hi.dim.coding} examines biophysical mechanisms
that restriction enzymes might use to attain high dimensional coding.
Section \ref{sec:coding.spaces}
explores the relationship between high dimensional codes and
restriction enzymes and
examines additional details of how restriction enzymes
recognize DNA in different dimensions.
In Section
\ref{sec:fitness.landscape}
we discuss how the coding space may represent the
biological fitness landscape through which restriction enzymes evolve.
Finally in Section
\ref{sec:transcription.factors}
we apply the theory to transcription factors in general
and find that they are probably functioning not only in
a high dimensional space, but the space dimensionality
is not an integer so (by definition) they function in 
a high dimensional fractal space. 

\section{A lower bound on the dimensionality of molecular machines}
\label{sec:lower}

Molecular machines are
molecules that select
specific states while dissipating energy
\cite{Schneider.ccmm}.
The information, in bits, that a molecular machine 
can gain is the base 2 logarithm of the number of states
it selects amongst
\cite{Schneider.Ehrenfeucht1986, Schneider.Stephens1990, Schneider.ev2000}.
The maximum number of bits that
can be gained for the energy dissipated
in a communications system is the channel capacity
\cite{Shannon1949}.
For molecular machines, we call the corresponding measure
the molecular machine capacity
\cite{Schneider.ccmm}.
Formulas for the capacities contain a term that represents
the dimensionality of the coding space in which the
state spheres are packed.
Therefore a rearrangement of the formula leads to an equation
for the dimensionality.
This provides a step towards
understanding the nature of the coding space of molecular machines.

The maximum number of distinct choices that a molecular machine can
make in the presence of thermal noise $N$ by dissipating energy $P$
depends on these two factors and also on the number of independent
moving parts of the machine or `pins,' $\ds$, following the lock and
key analogy of molecular machines~\cite{Schneider.ccmm}. 
In communications,
the channel capacity sets the upper bound on the rate
that information can be faithfully transmitted~\cite{Shannon1948, Shannon1949}.
Corresponding to the
channel capacity
of communications systems
a molecular machine's capacity is:
\begin{equation}
C = \ds \log_{2}{ \left( \frac{P}{N} + 1 \right) }
\;\;\;\;\; \mbox{(bits per operation)} ~,
\label{eqn.Cy}
\end{equation}%
where a molecular machine operation is, for example, the process of
going from non-specific to specific DNA binding by a nucleic acid
recognizer~\cite{Schneider.ccmm}.
This formula was derived by counting the maximum number of distinct
molecular states, represented as spheres in a high dimensional space
(see \cite{Schneider.nano2,Schneider2006,Schneider-brmit2010} for reviews),
assuming white Gaussian noise.
At the molecular level,
relevant to the functionality of
molecular machines such as restriction enzymes,
the noise is overwhelmingly thermal,
justifying the use of Shannon's results.
In order to faithfully transmit information,
Shannon's channel capacity theorem~\cite{Shannon1949} implies that the
sequence information a nucleic acid recognizing molecular machine uses
to locate its binding sites, $R = \rsequence$
\cite{Schneider.Ehrenfeucht1986}, can evolve up to but not beyond this
capacity:
\begin{equation}
R \le C ~.
\label{eqn.R.Cy}
\end{equation}
For nucleic acid recognizers,
$\rsequence$ is the area under a sequence logo
\cite{Schneider.Stephens1990}.
The dimensionality of the coding space used to describe these states is:
\begin{equation}
D = 2 \ds
\label{eqn.D.ds}
\end{equation}
since there are both a phase and an amplitude for each of the
independent oscillator pins that describe the motions of a molecule at
thermal equilibrium \cite{Schneider.ccmm}.
Combining equations~\eq{eqn.Cy},~\eq{eqn.R.Cy}, and~\eq{eqn.D.ds}
gives a lower bound for the dimensionality:
\begin{equation}
\frac{2 R}{\log_{2}{\left( \frac{P}{N} + 1 \right)}} \le D ~.
\label{eqn.lower.D.bound}
\end{equation}
This lower bound is a function of the information gain $R$ and the $P/N$ ratio.

\section{Applying the dimensional lower bound to restriction enzyme coding space}
\label{sec:apply.lower}

The maximum theoretical isothermal efficiency of a molecular machine
is defined entirely by the dissipated energy $P$ and the thermal noise
$N$ in terms of the normalized energy dissipation $\rho = P/N$:
\begin{equation}
\epsilon_t = \frac{\ln \left(\frac{P}{N} + 1 \right)}{\frac{P}{N}}
             =     \frac{\ln \left(\rho + 1 \right)}{\rho}
\label{eqn.epsilon.t}
\end{equation}
where $0 < \epsilon_t \le 1$
\cite{Schneider-emmgeo2010}.
This expression was first used to describe the efficiency of satellite
communications in terms of the `signal-to-noise' ratio, $P/N$
\cite{Pierce.Cutler1959}.

The efficiency of EcoRI and
other molecular machines is observed to be close to
$70$\%~\cite{Schneider-emmgeo2010}.  This can be explained if $P
\approx N$, in which case equation~\eq{eqn.epsilon.t} shows
$\epsilon_t \approx \ln 2 \approx 0.69$.
The relationship between $P$ and $N$
measures the distance between hyperspheres
(\fig{fig.packing})
so
$P=N$
implies that
the state of being bound to one sequence is distinct from the state of
being bound to a different sequence.

When there is a choice to be made among several molecular states, such
as the strong discrimination EcoRI makes between GAATTC and single
base changes of that sequence~\cite{Lesser.Jen-Jacobson1990}, then the
molecular machine operates under the condition that its states are
separated, which has been shown geometrically to be equivalent to
\begin{equation}
P > N
\label{eqn.P.N}
\end{equation}
(\fig{fig.packing})
\cite{Schneider-emmgeo2010}.
This inequality limits the efficiency to $70$\%.
Substituting into equation~\eq{eqn.lower.D.bound}, the inequality~\eq{eqn.P.N} implies that
\begin{equation}
2 R < D ~.
\label{eqn.R.D.noequal}
\end{equation}
For fully evolved bistate molecular machines the dimensionality of the coding space is
more than twice the information content of a binding site when the latter is expressed in bits.
Thus, a lower bound of the dimensionality for the EcoRI coding space is found by noting that GAATTC is $6$ bases or $12$ bits, so EcoRI operates in a coding space of at least $24$ dimensions.
Similarly, as a consequence of the inequality~\eq{eqn.P.N}, $\ds$
supplies a lower bound for the channel capacity in
equation~\eq{eqn.Cy},
\begin{equation}
\ds < C ~.
\label{eqn.ds.C.bound}
\end{equation}
For example, given $P>N$, if $P=31N$ then $C=5 \ds$, so $C > \ds$.

\section{An upper bound on the dimensionality of molecular
machines}\label{sec:upper}

The higher the dimension that a molecular machine can work in,
the more the probability density tightens around the radius
of the hyperspheres
\cite{Schneider.ccmm}.
This suggests that biological systems may tend to evolve to
extremely high dimensions to reduce the error rate caused by
switching between the hyperspheres.
So having determined
a lower bound on the dimensionality of a molecular machine
(equation \eq{eqn.lower.D.bound})
is tantalizing but unsatisfying because
biological systems may have much higher dimensionality.
For this reason we sought an upper bound on the dimensionality.

The dimensionality of a molecule is related to the number
of degrees of freedom ($\nu$) that a molecule has.
For $n$ atoms there are 3 independent
axes each atom can move on, but the three
translational motions and three rotations about the axes do not
contribute
to the functioning
of the machine, so there are only
\begin{equation}
\nu = 3n -6
\label{eqn.nu.n}
\end{equation}
degrees of freedom.
For water $n = 3$ so $\nu = 3$ normal modes
that can be observed in the vibrational spectrum of the molecule.
These three motions can be described by common arm exercises
with the head representing oxygen and the fists hydrogen:
pushup/pullup ($\nu_1$ symmetric stretch),
jumping jack ($\nu_2$ bending mode)
and
one-two punch ($\nu_3$ asymmetric stretch)
\cite{Chaplin2000}.

Although
the number of degrees of
freedom of an entire molecule consisting of $n$ atoms is $3n -6$,
the relevant number of degrees of freedom involved
in the molecular machine selection process coding space ($D$)
is most likely much smaller
\cite{Schneider.ccmm}:
\begin{equation}
D = 2 \ds \ll 2 \nu = 2(3n - 6) ~,
\label{eqn.ds.real.two}
\end{equation}
because to be able to evolve each molecular machine `pin'
consists of an average of up to $n/\ds \gg 1$ atoms.
For a large molecule like EcoRI with thousands of atoms,
the relevant degrees of freedom ($D$)
for DNA binding will be much smaller
than given by equation
\eq{eqn.ds.real.two},
so
that relationship does not give a useful upper bound.

As Jaynes pointed out~\cite{Jaynes1983,Jaynes1988},
based on the classical equipartition theorem,
the energy per degree of freedom of a single thermal oscillator
in a molecular machine (lock pin) is $\frac{1}{2}\kb T$;
with $D$ degrees of freedom the total thermal noise flowing through
a molecule during one dissipation step of $P$ that selects a
specific molecular state is
\begin{equation}
N = \frac{1}{2} \kb T \cdot D
\;\;\;\;\;\;\mbox{(joules per operation)}
\label{eqn.Ny.D}
\end{equation}
(see also equation (31) in~\cite{Schneider.ccmm}).

For molecules that make distinct decisions by selecting between
nonoverlapping hyperspherical states, the inequality~\eq{eqn.P.N}
applies.
Substituting \eq{eqn.P.N}
into equation~\eq{eqn.Ny.D},
\begin{equation}
D < \frac{P}{\frac{1}{2}\kb T} ~.
\label{eqn.first.upper.D.bound}
\end{equation}
This provides a upper bound on the functional dimensionality, whereas
equation~\eq{eqn.lower.D.bound} provides a lower bound.

We convert equation~\eq{eqn.first.upper.D.bound} to a more useful form by noting that the energy available in coding space  for making selections at one temperature and pressure~\cite{Schneider.ccmm,Schneider-emmgeo2010,Schneider-brmit2010} is the Gibbs free energy:
\begin{equation}
P = -\deltaGnaught
\; \; \; \; \; \mbox{(joules per operation)}
\label{eqn.Py.deltaGnaught}
\end{equation}
\cite{Schneider-emmgeo2010}.
The maximum number of bits that can be gained for that free energy dissipation is
\begin{equation}
\renergy \equiv - \deltaGnaught / \Emin
\; \; \; \; \; \mbox{(bits per operation)}
\label{eqn.Renergy}
\end{equation}
\cite{Schneider-emmgeo2010}.
$\Emin$ can be derived from information theory~\cite{Felker1952}
or the second law of thermodynamics~\cite{Szilard1929,Schneider.edmm}.
It serves as an ideal conversion factor between energy and bits:
\begin{equation}
\Emin = \kb T \ln 2
\; \; \; \; \; \mbox{(joules per bit)}
\label{eqn.Emin}
\end{equation}
\cite{Schneider-emmgeo2010}.
Further, a `real' isothermal efficiency
$\epsilon_r$, that may be less than the theoretical
efficiency of equation \eq{eqn.epsilon.t},
\begin{equation}
\epsilon_r \le \epsilon_t ~,
\label{eqn.epsilon.r.le.epsilon.t}
\end{equation}
can be measured
by the information gained, $R$, \emph{versus} the information that could have been gained for the given energy dissipation, $\renergy$:
\begin{equation}
\epsilon_r = R/\renergy
\label{eqn.epsilon}
\end{equation}
\cite{Schneider-emmgeo2010}.
Successively combining
equations~\eq{eqn.Py.deltaGnaught} to \eq{eqn.epsilon} gives
\begin{eqnarray}
P &=& \Emin \renergy \\ \nonumber
  &=& (\kb T \ln 2) \, \renergy \\ \nonumber
  &=& \kb T R \ln 2 /\epsilon_r ~. %
\label{eqn.Py.R}
\end{eqnarray}
Inserting this result into equation~\eq{eqn.first.upper.D.bound} gives
\begin{equation}
D < \frac {2 R \ln 2 } {\epsilon_r} ~,
\label{eqn.upper.D.bound}
\end{equation}
which we recognize as an upper bound on the coding space dimensionality as a function of the information gain $R$ and the isothermal efficiency $\epsilon_r$.

\section{Pincers on the dimensionality of molecular machines}
\label{sec:pincers}

Having determined both
a lower bound
(equation \eq{eqn.lower.D.bound})
and an upper bound
(equation \ref{eqn.upper.D.bound})
on the dimensionality of the coding space, 
we have the opportunity to determine what will happen
as the molecular machine evolves to be optimally efficient.

Combining equations~\eq{eqn.lower.D.bound} and~\eq{eqn.upper.D.bound} gives
\begin{equation}
\frac{2 R \ln 2}{\ln{\left(\rho + 1 \right)}} \le D < \frac{2 R \ln 2}{\epsilon_r} ~,
\label{eqn.lower.upper.D.bound}
\end{equation}
where we have recast the left hand side in terms of $\rho = P / N$ and expressed the logarithm in base $e$ to emphasize the striking symmetry of the two sides
(the numerators are identical and the denominators are
related by equations \eq{eqn.epsilon.t} and \eq{eqn.epsilon.r.le.epsilon.t}).
The dimensionality is constrained to lie between these bounds, which form closing `pincers' as the molecular machine evolves to become optimal.
In the limit as
$P \rightarrow N$
and
$\epsilon_r$ evolves to its maximum value of $\ln 2$~\cite{Schneider-emmgeo2010},
both sides converge to $2 R$, and $D$ is squeezed between them.
An optimally evolved molecular machine will operate in $2R$ dimensions:
\begin{equation}
D_{\text{optimal}} = 2 R ~.
\label{eqn.D.optimal.bits}
\end{equation}

However, because the state hyperspheres have a finite
thickness~\cite{Schneider.ccmm}, $P$ must at least slightly exceed
$N$, and the left hand side
of \eq{eqn.lower.upper.D.bound}
remains slightly smaller than $2 R$. 
Likewise, because of equations~\eq{eqn.epsilon.t}
and~\eq{eqn.epsilon.r.le.epsilon.t}, $P > N$ also means that the
efficiency $\epsilon_r$ is slightly below $\ln 2$, which makes the
right hand side
of \eq{eqn.lower.upper.D.bound}
sightly larger than $2 R$.  So the dimensionality is
restricted to a small interval.  This allowed variation in the coding
space dimensionality is caused by the effective thickness of the
sphere surface fuzziness (which depends on the dimensionality $D$
itself) and the evolutionarily acceptable error rate determined by the
environment which limits how closely the coding spheres can approach
each other and still allow survival~\cite{Schneider.ccmm}.

For example, DNA polymerase has a certain error rate,
and of course if that error rate were to increase
the organism would experience a higher number mutations and
be at a selective disadvantage.
However,
there are also
mutations of the polymerase that decrease its error
rate~\cite{Herr.Preston2011}.
This would reduce the mutation load on the organism, but presumably it
does not occur in the wild because the organism would then be less
able to evolutionarily adapt to changing conditions compared to
siblings that have the higher error rate.
In many but not all cases they would also replicate more slowly. 
Likewise, the error rate for translation is about $1$ in $1000$ amino
acids
\cite{Wohlgemuth.Rodnina2010}
which means that roughly one in every three proteins has an error.
Yet organisms survive quite well at this error rate.
The error rate set by the environment of the organism in turn determines the acceptable placement and thickness of the spheres.

\begin{figure}[tbhp]%
\begin{center}
\scalebox{1.8}{\includegraphics*{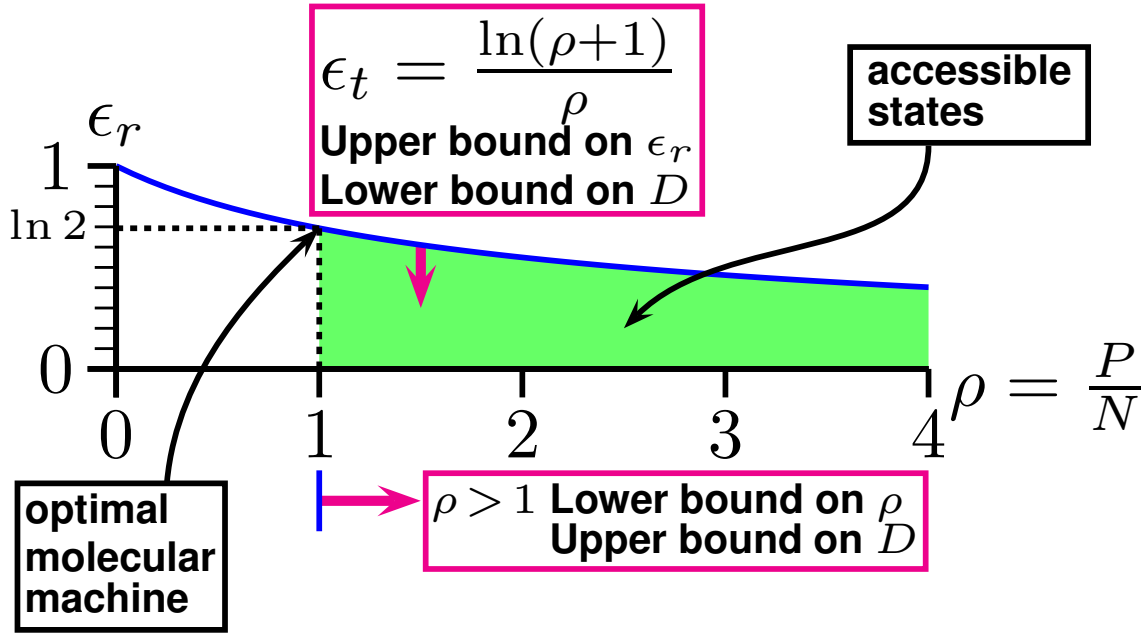}} %
\end{center}
\caption{Isothermal efficiency curve
for molecular machines showing
bounds that constrain
the coding space dimensionality $D$.
Real
molecular machines that select between two or more distinct states
may have parameters anywhere in the shaded (green) area
in which the real isothermal efficiency $\epsilon_r$
is bounded above
by the theoretical isothermal efficiency $\epsilon_t$
(equation \eq{eqn.epsilon.t})
and to the left
by the power to noise ratio $\rho = P/N > 1$ (equation \eq{eqn.P.N}).
During evolution,
they tend to lose unnecessary energy dissipation,
which decreases $\rho$ towards the lower
limit of $\rho=1$.
Independently,
they tend to increase their information use ($R$) for the energy
dissipated,
which increases $\epsilon_r$ toward
the theoretical maximum $\epsilon_t$
determined by the channel capacity.
These factors lead to an `optimal' molecular machine in which
$\rho = 1$
and
$\epsilon_r =\epsilon_t = \ln 2$.
At that point the dimensionality
has been squeezed in a pincers (equation \eq{eqn.lower.upper.D.bound})
until it reaches $D = 2 R$.
}
\label{fig.effcurvebounds}
\end{figure}

Note that the lower bound constraint comes from equation~\eq{eqn.R.Cy},
the channel capacity theorem of Shannon~\cite{Shannon1949}, which limits the efficiency $\epsilon_r$.
The upper bound comes from equation~\eq{eqn.P.N}, the finite energy available to perform state selections ($P$) relative to the thermal noise ($N$), which satisfies the biologically required separation of molecular states~\cite{Schneider-emmgeo2010}.
These two independent bounds are plotted on a graph of the efficiency curve
(\fig{fig.effcurvebounds},
\figmargin{fig.effcurvebounds}
equation~\eq{eqn.epsilon.t}~\cite{Schneider-emmgeo2010}).

The efficiency curve is an upper bound representing functioning at the
channel capacity (equation~\eq{eqn.R.Cy}).  Points below the curve
have $R < C$.  Since the dimensionality parameter $D = 2\ds$ is part
of the upper bound for $C$ in equation~\eq{eqn.Cy}, this leads to the
lower bound on $D$ in equation~\eq{eqn.lower.D.bound}.  Independently
of that, the $\rho = P/N > 1$ ratio on the horizontal axis
of~\fig{fig.effcurvebounds} is orthogonal to the efficiency and
channel capacity of the vertical axis.  The thermal noise $N$ is
determined by the absolute temperature of the molecular machine and
the dimensionality (equation~\eq{eqn.Ny.D}).  Since $P$ is an upper
bound on the noise, equation~\eq{eqn.P.N} leads to an upper bound on
the dimensionality in equation~\eq{eqn.upper.D.bound}.

These two independent constraints on $\epsilon_r$ and $\rho$ determine
the possible range of the dimensionality.  As shown
previously~\cite{Schneider-emmgeo2010}, the normalized energy
dissipation $\rho$ will tend to decrease over evolutionary time
because excess contacts that are not required for maintaining
information will be lost by mutation.  This decrease will continue
until at $\rho = 1$ the distinctness of molecular states is threatened
by a large error rate caused by the increasing intersection of after
state spheres.  Meanwhile, the efficiency $\epsilon_r$ will tend to
increase up to $\ln 2$, squeezing the dimensionality towards a single
value, $2 R$ by equation~\eq{eqn.lower.upper.D.bound}.  This result is
consistent with the dimensional analysis of
Collier~\cite{Collier2017}, who showed that information is related to
the degrees of freedom, which is of course the dimensionality of the
space.

\section{Dimensionality of restriction enzyme coding space}
\label{sec:dimensionality}

Now that we know that the dimensionality of an optimal
molecular machine is simply twice the number of bits that it
selects amongst
(equation \eq{eqn.D.optimal.bits})
we can determine the dimensionalities of
the thousands of known restriction enzymes
if we assume that they too are optimal.
In the case of restriction enzyme EcoRI,
experimental data \cite{Clore.Davies1982} show that
the efficiency is close to
$\epsilon_r = \ln 2$~\cite{Schneider-emmgeo2010}, so that $\rho = P/N$
must be close to $1$.  
Thus, although complete data are not available
about specific and non-specific binding of other restriction enzymes,
we assume that, like EcoRI,
these molecular machines
have also evolved to be close to $\ln 2$ efficiency
as shown in \fig{fig.effcurvebounds}.
We believe this is a reasonable assumption
because they should all distinguish their binding sites from
other sequences with as small an error as possible so
their hypersphere states should not overlap significantly ($P>N$),
and they should maximize the information gain for energy dissipation
(the efficiency) but
they cannot exceed the efficiency upper bound 
shown in equation \eq{eqn.epsilon.t}
\cite{Schneider-emmgeo2010, Schneider-brmit2010}.

Since a $6$ base cutting restriction enzyme
recognizes $R = 6 \times 2\ \mathrm{bits} = 12$ bits, from
equation~\eq{eqn.lower.upper.D.bound}, we see that the EcoRI coding
space must be close to $24$ dimensions.
If we characterize a restriction enzyme by the number of bases it recognizes,
then there are a maximum of two bits per base,
so the dimensionality of a $70$\% efficient molecule is:
\begin{equation}
D = 4 \times \mbox{(bases per binding site)} ~.
\label{eqn.D.bases}
\end{equation}
Thus, EcoRI, which has a $6$ base recognition site GAATTC,
works in $24$ dimensions, while TaqI, which recognizes only the 4
bases TCGA, should work in $16$ dimensions.  The highest known
dimension used by a restriction enzyme is $32$ dimensions for
restriction enzymes such as NotI (GCGGCCGC)
and SfiI (GGCCNNNNNGGCC), which cut DNA at patterns $8$ base pairs
long~\cite{Qiang.Schildkraut1987,Roberts.Macelis2015}.

In the case of restriction enzymes that digest at partially variable
patterns such as GT(T/C)(A/G)AC (HincII), we can use the information
needed to describe the pattern to predict the dimension.
In this case, for the first two and last two bases (GT and AC), a
total of $8$ bits are required, while for each of the middle two bases
only $1$ bit is required to distinguish two of the four bases so $R =
10$ bits per site~\cite{Schneider.Ehrenfeucht1986}, even though the
binding site is $6$ bases long.
Thus, if HincII is optimal at $70$\% efficiency, it should operate in
$2R=20$ dimensions.
In the special case where a base is avoided by a restriction enzyme,
we record the information as $2 - \log_2 3$ bits for that
base~\cite{Schneider.Ehrenfeucht1986}.  So formula~\eq{eqn.D.bases}
has only a limited application.  The number of bits in a binding site
is not strictly computed from the physical length of the site, but
rather from the average number of bases if all the information were
compressed into the smallest region
possible~\cite{Schneider.Ehrenfeucht1986}.  This is the `area' under a
sequence logo~\cite{Schneider.Stephens1990}.

\begin{figure}[tbhp]%
\centering %
\scalebox{0.70}{\includegraphics*{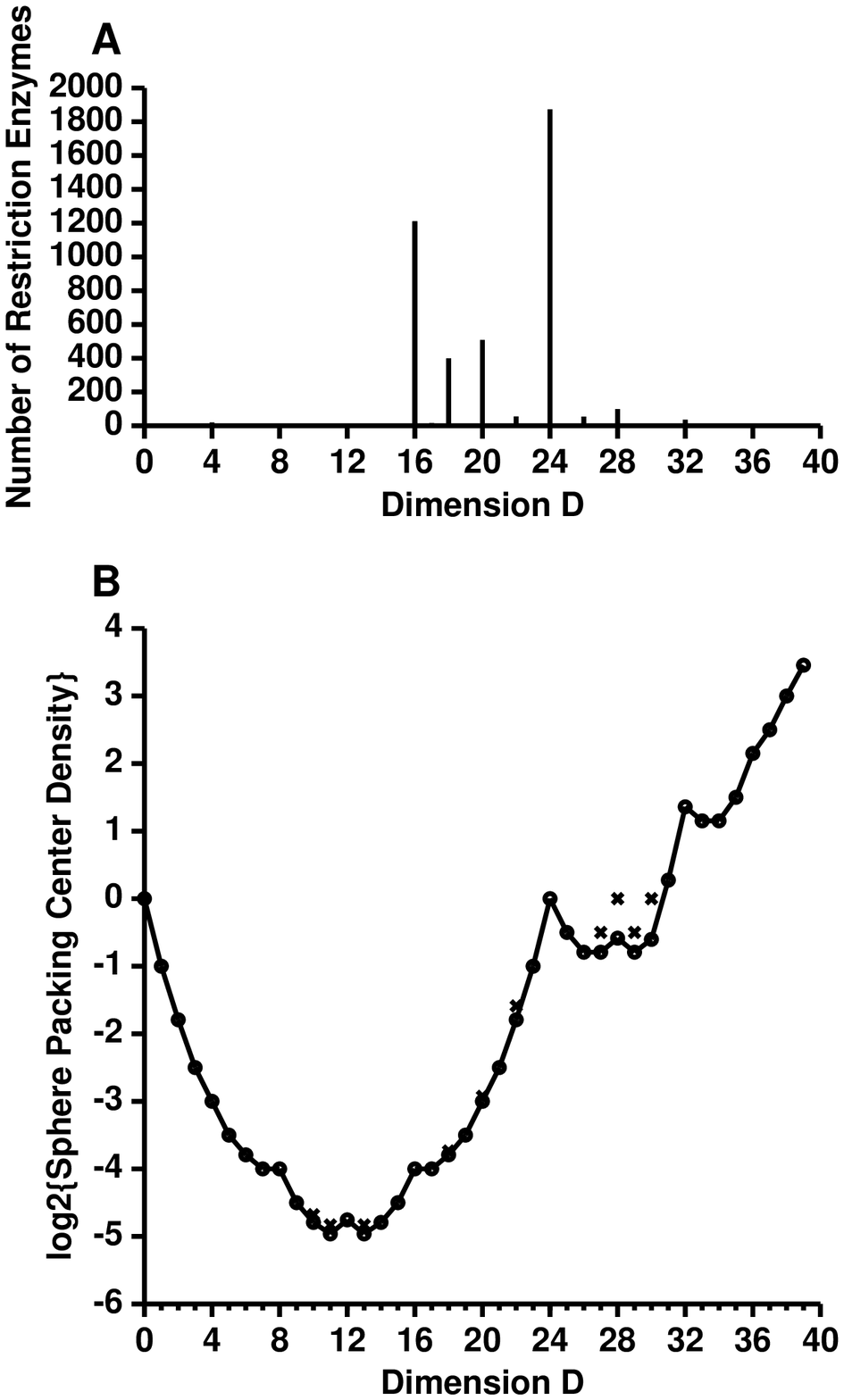}} %
\caption{Comparison of restriction enzyme frequency and
best known sphere packing density in different dimensions.
A. Coding dimensions used by restriction enzymes.
The number of enzymes at each dimensionality is plotted
from~\tablereference{enzymetable}.
B. Best known sphere packings in high dimensions
were given by
Conway and Sloane~\cite{Conway.Sloane1998,Sloane1998}.
The graph is equivalent to their Figure~1.5;
see
Table I.1(a), Table I.1(b) on
pages xix and xx; and pages 14 to 16.
The updated sphere center density formulas used here
were from %
\url{http://www.math.rwth-aachen.de/~Gabriele.Nebe/LATTICES/density.html} %
(Last modified Feb. 2012, accessed Jan 06, 2018).
The sphere center density,
$\delta$,
is the number of sphere centers per unit volume when sphere radii are
set to 1.
Without the logarithm, a graph of $\delta$ versus $D$
appears nearly flat from
$D = 7$
to
$D = 18$.
Circles ($\circ$) represent lattice packings;
x's ($\times$) represent nonlattice packings.}
\label{fig.restpack}
\end{figure}

The predicted dimensionality of over $4000$
Type II restriction enzymes in
Roberts' REBASE database \cite{Roberts.Macelis2015} is given
in~\tablereference{enzymetable}  \tablemargin{enzymetable}
and~\fig{fig.restpack}A.
\figmargin{fig.restpack}
There are two major peaks at $24$ and $16$ dimensions, corresponding
to $6$ and $4$ base cutters.  There is also a minor peak at $32$
dimensions for the $8$ base cutters.

The 4297 enzymes we studied consist mostly (98\%)
of Type II restriction enzymes.
A set of 482 Type I enzymes,
which cut randomly at a distance away from where they bind specifically,
were also collected from REBASE
\cite{Roberts.Macelis2015}
\\
(\url{http://rebase.neb.com/rebase/rebadvsearch.html}
on 9/30/2019
using the search paramters:
`Type I',
`Specificity subunit',
`Non-putative',
`Prototype'
and columns:
'Enzyme name' and
'Recognition sequence').
They
primarily use
24 (105 cases),
26 (112 cases) and
28 (206 cases) dimensions.
Since the additional enzymes are a small fraction of the database
our conclusions would not change by including them.

\begin{table}
\center
\colorlet{tablerowcolor}{gray!19.0} %
\rowcolors{4}{white}{tablerowcolor} %
\scalebox{0.90}{%
\begin{tabular}{llcrrrl}
Example & Sequence & Compressed & Bits & Dimension & Number \\
Restriction &   & Bases, $\lambda=R/2$ & $R$ & $D=2R$ & $N$ \\
Enzyme &   &   & (pins) &   &   \\
\hline
AbaSI & C(11/9) & 1.00 & 2.00 & 4.00 & 20 \\
MspJI & CNNR(9/13) & 1.50 & 3.00 & 6.00 & 1 \\
RlaI & VCW & 1.71 & 3.42 & 6.83 & 1 \\
SgeI & CNNGNNNNNNNNN$\downarrow$ & 2.00 & 4.00 & 8.00 & 6 \\
AspBHI & YSCNS(8/12) & 2.50 & 5.00 & 10.00 & 1 \\
PsuGI & BBCGD & 2.62 & 5.25 & 10.49 & 1 \\
SgrTI & CCDS(10/14) & 2.71 & 5.42 & 10.83 & 2 \\
CviJI & RG$\downarrow$CY & 3.00 & 6.00 & 12.00 & 9 \\
LpnPI & CCDG(10/14) & 3.21 & 6.42 & 12.83 & 1 \\
EcoBLMcrX & RCSRC(-3/-2) & 3.50 & 7.00 & 14.00 & 1 \\
M.NgoDCXV & GCCHR & 3.71 & 7.42 & 14.83 & 1 \\
TaqI & T$\downarrow$CGA & 4.00 & 8.00 & 16.00 & 1210 \\
Bsp1286I & GDGCH$\downarrow$C & 4.42 & 8.83 & 17.66 & 16 \\
AvaII & G$\downarrow$GWCC & 4.50 & 9.00 & 18.00 & 396 \\
Pin17FIII & GGYGAB & 4.71 & 9.42 & 18.83 & 2 \\
HincII & GTY$\downarrow$RAC & 5.00 & 10.00 & 20.00 & 507 \\
Cco14983V & GGGTDA & 5.21 & 10.42 & 20.83 & 1 \\
PpuMI & RG$\downarrow$GWCCY & 5.50 & 11.00 & 22.00 & 55 \\
EcoRI & G$\downarrow$AATTC & 6.00 & 12.00 & 24.00 & 1864 \\
Rba2021I & CACGAGH & 6.21 & 12.42 & 24.83 & 10 \\
PspXI & VC$\downarrow$TCGAGB & 6.42 & 12.83 & 25.66 & 1 \\
RsrII & CG$\downarrow$GWCCG & 6.50 & 13.00 & 26.00 & 54 \\
SgrAI & CR$\downarrow$CCGGYG & 7.00 & 14.00 & 28.00 & 99 \\
KpnBI & CAAANNNNNNRTCA & 7.50 & 15.00 & 30.00 & 2 \\
SfiI & GGCCNNNN$\downarrow$NGGCC & 8.00 & 16.00 & 32.00 & 36 \\
\end{tabular}

}
\caption{
Coding space dimensionality ($D$)
and number ($N$) of restriction enzymes.
The information content in bits, $R$, of the recognition sequence of
4297 %
restriction enzymes from
REBASE (restriction enzyme database)
\url{http://rebase.neb.com} or
\url{ftp://ftp.neb.com/pub/rebase/} version allenz.801 
(Dec 27 2017)~\cite{Roberts.Macelis2015}
was computed.
A fully conserved base
(A, C, G, T)
contributes $2 - \log_2 1=2$ bits,
two possibilities
(%
R=G/A,
Y=C/T,
M=A/C,
K=G/T,
S=C/G,
W=A/T)
contributes $2 - \log_2 2=1$ bit,
three possibilities
(%
B=C/G/T,
D=A/G/T,
H=A/C/T,
V=A/C/G)
contributes $2 - \log_2 3 \approx 0.42$ bits
and
any allowed base
(N)
contributes $2 - \log_2 4 = 0$
bits~\cite{Dixon.Stockwell1985,Schneider.Ehrenfeucht1986}.
The sum of the information at each base, $R$, was used to find the
corresponding
number of compressed bases ($\lambda=R/2$)
and then
the coding dimension ($D=2R$),
assuming that each enzyme has an efficiency of $\epsilon_r=\ln 2$ and
$\rho = 1$ so that there is a unique dimension
according to equation \eq{eqn.lower.upper.D.bound}.
The most commercially available
enzymes and their reported recognition sequences are given
as examples.
When the DNA backbone cleavage site is known it
is indicated by an arrow ($\downarrow$).
The distance to cleavage sites outside the given sequence is
shown in parenthesis for the corresponding and complementary strands.
Star activity (variation within the canonical site)
and flanking sequence effects are found for many restriction
enzymes~\cite{Kamps-Hughes.Johnson2013}.
However,
the patterns in the database
are reported as consensus sequences
that may distort the information
content~\cite{Schneider.Zen2002},
and so may affect the results given here.
}
\label{enzymetable}
\end{table}

\section{Biological lattices in high dimensional spaces
explain restriction enzyme 4 and 6 base pair preferences}
\label{sec:lattices}

Ever since Shannon published his
theory of hypersphere packing as a description of a communications system
\cite{Shannon1949}
mathematicians and
engineers have been determining how best to pack spheres together
in high dimensional spaces
\cite{Conway.Sloane1998}.
The restriction enzymes appear to favor particular dimensions
for their coding spaces,
so we can compare their preferences to the best known packings
that humans have determined.

In two dimensions there are two ways to regularly
pack circles: in a square lattice or in a hexagonal lattice
(\fig{fig.packing}).
The square packing fills
$\Delta = \pi r^2 / (2 \times r)^2 = 79$\%
of the plane, while a hexagonal packing fills
$\Delta = \pi / \sqrt{12} = 91$\%~\cite{Conway.Sloane1998}.
Hexagonal packing is more dense than square packing.
In general, the sphere packing density is
\begin{equation}
\Delta = \frac{V_n r^n}{(\mbox{\scriptsize det}\, \Lambda)^{1/2}}
\label{eqn.bigDelta}
\end{equation}
where $V_n$ is the volume of an $n$ dimensional
sphere
with radius $r$
\cite{Sommerville1929,Kendall1961,Hayes2011n},
and
$\mbox{det}\, \Lambda$ is the determinant of the lattice $\Lambda$.
The determinant provides the volume of the polytope
that the sphere is encased in.
For convenience,
Leech introduced the concept of 
the sphere center density
\begin{equation}
\delta = \frac{\Delta}{V_n}
\label{eqn.littledelta}
\end{equation}
which counts the average number of sphere centers per unit volume of the
space~\cite{Leech1964}.
Leech, Conway and Sloane rescaled or normalized the sphere center
density in several different ways to
emphasize the symmetries
of the sphere center density
as a function of dimension
\cite{Leech1964, Conway.Sloane1982, Conway.Sloane1998}.
As these rescalings do not have any biological significance
that we are aware of, we only take the logarithm of $\delta$
to graph the best known sphere packings up to $40$ dimensions
(\fig{fig.restpack}B).
Since each sphere center represents one sphere,
and
in a biological context
spheres represent biological states,
the center density is a measure of the number of states
available to the system.

Shannon recognized that the packing of spheres in higher dimensions
corresponds to the problem of faithfully transmitting a series of
distinct messages over a noisy communications
channel~\cite{Shannon1949,Sloane1984,Conway.Sloane1998,Cipra1990}. 
In this model,
transmitted messages are points in a high dimensional space, and
each sphere represents a message received with
Gaussian noise added along each dimension.
Most of the sphere density is on the surface in high
dimensions~\cite{Schneider.ccmm}.
To avoid message ambiguity, the spheres must not intersect, which
spaces the transmitted messages
and allows a decoding that removes
the noise from the received signal.
The total volume available in which to pack spheres is
a large sphere whose radius is
determined by the power and thermal noise absorbed by the receiver,
while the volume of a smaller message sphere
is determined by the thermal noise alone~\cite{Shannon1949}.

A corresponding theory also describes the states of molecular
machines as spheres in high dimensional space
\cite{Schneider.ccmm}.
In both theories, the maximum number of possible messages
(or molecular states), known as the capacity,
is determined from the number of small spheres
that can be packed together inside the larger sphere.
As in $2$ dimensions, there are many possible ways to pack high
dimensional spheres; the more spheres that can be packed together,
the more the channel can be utilized.
Because of its application to communications
and a variety of mathematical and physics fields,
the highest sphere packing densities in various dimensions have
been determined, as shown in
\fig{fig.restpack}B.

The histogram
in \fig{fig.restpack}A
is compiled from data corresponding
to $4297$ restriction enzymes, whereas the center density
graph in \fig{fig.restpack}B
is an exact
mathematical result for $51$ lattice packings.  It is striking how
well they match at $D=16$ and $D=24$, corresponding to $4$ and $6$ base
cutters.  The histogram also has minor peaks at $D=18$ and $D=20$ that are
not reflected in the density plot.  We argue below that this happens
for biological reasons.  Similarly, the peak at $D=32$ in the center
density plot does not have a corresponding large peak in the histogram.
As discussed below,
we explain that this is for biological reasons as evolution seems to
prefer recognition and excision of smaller pieces of DNA.

The most dense known sphere packing is in $24$ dimensions, a
packing known as the Leech lattice (symbolized as $\Lambda_{24}$)
\cite{Leech1964,Conway.Sloane1998,
Stewart2003,    %
Cohn.Elkies2003, %
Cohn.Kumar2004, %
Cohn.Kumar2009, %
Klarreich2016,
Viazovska2017, Cohn.Viazovska2017}.
This packing has been extensively studied,
and because of its density it
was used in a commercial Motorola modem
\cite{Lang.Longstaff1989}.
Surprisingly, the most common dimensionality of the restriction
enzymes is also $24$ dimensions, as shown in~\fig{fig.restpack}A.
This suggests that EcoRI and the other restriction enzymes
may be $6$ base cutters because that takes advantage of the
dense packing of the Leech lattice.
In other words, we hypothesize that
the reason so many restriction enzymes are $6$ base
cutters is that they have discovered the Leech lattice packing
by Darwinian evolution.
How the Leech lattice is implemented by the atomic structure of
restriction enzyme proteins is not known.

For restriction enzymes,
the next most commonly used dimensionality
is $16$ dimensions (\fig{fig.restpack}A),
and we see in \fig{fig.restpack}B that
a good packing,
the Barnes--Wall lattice $\Lambda_{16}$
(BW${}_{16}$),
has also been found for this dimension
relative to the other dimensions of similar magnitude \cite{Sloane1998}.
There are an estimated $10^7$
good packings equivalent to
BW${}_{16}$
\cite{Conway.Sloane1995}.
Thus, the $4$ base cutting restriction enzymes may be using $8$ bit
recognition to take advantage of the good hypersphere
packings possible in $16$ dimensions.

The longest known restriction enzyme sites have $8$ bases,
and so these enzymes should use a $8 \times 4 = 32$ dimensional space.
Correspondingly,
$32$ dimensions also represents a peak in the known dense lattices
called the
Quebbemann's lattice, $Q_{32}$~\cite{Sloane1998}.
Cohn and Elkies report a peak in $28$ dimensions
\cite{Cohn.Elkies2003} and,
intriguingly, this corresponds to $99$ cases of $7$ base cutters
(\tablereference{enzymetable}).
Transcription factors in \emph{E.\ coli}
have information contents in the range $16$ to $23$
bits~\cite{Schneider.Ehrenfeucht1986};
these may function with the high density packings known
to exist above $40$ dimensions.

On the lower dimensional end,
it is worth noting that
there is a small local maximum for the density of
sphere packings at $12$ dimensions.
This would correspond
to a $3$ base long biological object
for which
the obvious candidate is the codon of the genetic code.
It may be that the genetic code functions in a $12$ dimensional space,
but the coding is probably not performed via the
cubic lattice $Z^{12}$
suggested by Sadegh--Zadeh
\cite{SadeghZadeh2000} %
since there are better packings such
as the Coxeter--Todd lattice
$K_{12}$~\cite{Conway.Sloane1998}.
Finally, in $8$ dimensions the most efficient possible sphere packing
is on an $E_8$ lattice~\cite{Viazovska2017};
this corresponds to $2$ base pair recognition.
Biologically this code might be used for precisely recognizing
and methylating CpG base pairs, the basis of an important
epigenetic control~\cite{Miranda.Jones2007, %
Song.Patel2012}.  %
Thus, all of the peaks in
\fig{fig.restpack}b
could correspond to known biological systems.

Restriction enzymes can evolve from one
dimension to the next since this only requires increasing or decreasing
the number of base contacts \cite{Pingoud.Pingoud2005,Chinen.Kobayashi2000}.
So there is some fluidity in the dimensions chosen, but
for several reasons
we do not expect a complete correspondence between \fig{fig.restpack}A
and \fig{fig.restpack}B.
First, because restriction enzymes have evolved, there
is a good deal of history in the current choices and some
of this may be locked in.
Some patterns will be common
simply because that particular bacterial species
is prevalent and their restriction enzymes were discovered more easily
than others.
Second,
the $8$ base enzymes ($32$ dimensions)
won't attack an invading DNA as frequently as shorter ones,
so bacteria may tend to avoid using higher dimensions. 
Third, short patterns that cut frequently 
would necessitate more self-protective methylation
and so would be expensive \cite{Chinen.Kobayashi2000}.
Fourth, evolutionary pressures to increase
or decrease the binding site information may not be equal
\cite{Chinen.Kobayashi2000}.
Finally,
unknown effects could come into play to eliminate, for example,
most of the
$5.5$ base ($22$ dimensional)
restriction enzymes
even though the
$5$ ($20$ dimensional)
and
$6$ ($24$ dimensional)
base sites are quite common.

The information content
of transcription factor
DNA binding sites
evolves based predominantly on the size of the genome
and the number of binding
sites~\cite{Schneider.Ehrenfeucht1986, Schneider.ev2000}.
Unlike transcription factors,
the information content of restriction enzyme sites
cannot evolve based on invading genomes
because there are no regular specific sequences to bind to.
However, the size of the intruding genome does provide some criterion
since
restriction enzymes protect bacteria from invading bacteriophage.
Typical sizes are on the order of $40,000$ base pairs, such as
$\lambda$ ($48502$ base pairs) \cite{Sanger.Petersen1982}
and
T7 ($39937$ base pairs) \cite{Dunn.Studier1983}.
The restriction enzyme must cut the invader
at least once, and preferably more, to disable the phage
genome.  Thus, it requires approximately 
$\log_2 48502 = 16$ bits in the site to attack $\lambda$ once.
A $12$ bit (6 base pairs) site such as EcoRI would cut $\lambda$
$2^{16 - 12} = 16$ times; $5$ sites are observed.
Perhaps this number is lower than expected because phage
evolve away from restriction sites;
EcoRI would cut T7 $8$ times but none are observed.
An $8$ bit ($4$ base pairs) site would cut more frequently
than a $12$ bit site
but the cell would
then have to methylate
$2^{12-8} = 16$ times as many sites.
Perhaps this is one reason that 6 base restriction
sites are more abundant than 4 base cutters:
6 bases is short enough that
phage are killed but also sufficiently long
that methylation is minimized.

Only the best choices of sphere packings
in biologically useful dimensions
may be reflected in the restriction enzymes.
The central suggestion of this paper is that,
although restriction enzymes are
highly divergent
\cite{Chmiel.Skowronek2005,
Bujnicki2000,
Bujnicki2003,
Gupta.Sharma2012,
Jeltsch.Pingoud1995},
most of them
have discovered that sphere packing in 16 and 24 dimensions
is more dense than packings in other dimensions.
This provides an explanation for why 4 and 6 base cutters have been found so
frequently.
In addition, the significant
peaks at $18$ and $20$ dimensions
in \fig{fig.restpack}A
suggest the biological use of dense codes in those dimensions
that may be consistent with known packings.

We regard an explanation as being a theoretical justification for an
observed feature within a data set.  Alternative reasons for
restriction enzyme 4 and 6 base preferences might including ease of
coding or efficacy of protein folding, but there needs to be a
specific hypothesis justified by a mathematical model in order to
provide a competing explanation.  To our knowledge, there isn't an
alternative model that explains the features we have noted.  In our
view, since restriction enzymes are highly divergent
\cite{Chmiel.Skowronek2005, Bujnicki2000, Bujnicki2003,
Gupta.Sharma2012, Jeltsch.Pingoud1995}, ease of coding and protein
folding are unlikely to explain the convergence to 4 and 6 bases. 
Also, our model provides a compelling explanation for which dimensions
are preferred.

\section{Mechanism of high dimensional coding}
\label{sec:hi.dim.coding}

For an evolved molecular machine
$D = 2 \ds$ (from equation \eq{eqn.D.ds})
and
$D = 2 R$ (from equation \eq{eqn.lower.upper.D.bound})
so
$\ds = R$.
So, curiously,
for an optimal molecular machine the number of bits
is the number of pins.
However, how the independent pins are implemented
in molecular architecture is a difficult open problem.
As in genetics, the underlying mechanism
of
DNA recombination
was not initially known
but the results,
linearity of genes,
were still valid.
Here, we know the dimensionality from the theory, but
we would also like to know how the molecule works.

There are at least two basic mechanisms
by which high dimensional coding could be implemented by molecules:
direct contacts and vibrational modes.
For example,
EcoRI cuts double stranded DNA at the sequence
GAATTC.
In the co-crystal between EcoRI and this sequence,
McClarin \emph{et al.}~\cite{McClarin.Rosenberg1986,Kim.Rosenberg1990}
observed that
each of the $6$ bases is contacted with two hydrogen bonds,
for a total of $12$ specific hydrogen bonds.
If each hydrogen bond
corresponds to a single `pin' of the molecular machine,
with two degrees of freedom per pin~\cite{Schneider.ccmm},
there would be $24$ dimensions.
That such contacts often act
independently,
and so could be coding space dimensions, is suggested by experiments
on several other
recognizers~\cite{Childs1985,Stormo.Gold1986,Barrick.Stormo1994,%
Takeda.Rivera1989,Lehming.MullerHill1990,Schneider.ccmm}.
However, experiments with mutant
EcoRI imply that it uses more than
just hydrogen bonding in
recognition~\cite{Heitman.Model1990b}, %
and bases of DNA recognition proteins
are not entirely independent~\cite{Man.Stormo2004, Bulyk.Church2002}.
Though including dinucleotides may be
sufficient~\cite{Stormo2011, Zhao.Stormo2012},
finding the important independent dimensions may be challenging.
All such pairwise correlations can be displayed with a
$3$ dimensional sequence logo~\cite{Bindewald.Shapiro2006}.
Alternatively,
the coding space could consist
of normal modes of molecular vibration since these
are by definition independent~\cite{Doruker.Kurkcuoglu2006}. %
In particular, localized vibrational modes called `discrete
breathers'~\cite{Csermely.Nussinov2010} %
may represent the molecular machine pins.

\section{Coding spaces}
\label{sec:coding.spaces}

In classical information theory, a continuous communications
signal, such as a song, can be represented by a series of independent
numerical values~\cite{Shannon1949}.  An
analog signal of duration $t$ seconds that has a range of frequencies
(bandwidth) $W$
is described by $D = 2 t W$ Fourier components.
Since these sine wave amplitudes are independent,
they define $D$ numbers and hence a single point in a $D$ dimensional
coding space.
Because they are designed from scratch,
the dimensionality in communications systems is known \emph{a priori}. 
By contrast molecular systems, which also have been shown to use
coding spaces~\cite{Schneider-emmgeo2010}, do not have a known
dimensionality so determining this parameter is an important step towards
fully characterizing and understanding their function.

Equations~\eq{eqn.lower.D.bound} and~\eq{eqn.upper.D.bound}
establish
lower and upper bounds on the dimensionality
of molecular machines.
These constraints can be represented geometrically
(\fig{fig.effcurvebounds}).
The restriction that the
information $R$ cannot be larger than the machine capacity
$C$
(equation \eq{eqn.R.Cy})
ultimately comes from Shannon's 1949 model
of communication in which he divided the volume
of a large `\emph{before}' sphere, representing the space of all possible
messages, by the volume of a small `\emph{after}' sphere,
representing a single message
expanded in all possible directions by thermal noise,
to determine
the maximum number of possible distinct messages $M$
in time $t$
and hence
the channel capacity in bits,
\begin{equation}
C = \lim_{t \rightarrow \infty}\frac{\log_2 M}{t}
\label{eqn.C.limit}
\end{equation}
\cite{Shannon1949}.
The corresponding model for molecular states
(equation \eq{eqn.Cy})~\cite{Schneider.ccmm}
leads to the
lower bound on the dimensionality
(equation~\eq{eqn.lower.D.bound}).
In this case there are two geometrical constraints, state spheres must
not intersect and the state spheres are confined to the larger sphere
defined by the available energy.
The observation of $70$\% efficient molecular machines comes from
the restriction that for biological states to be distinct, the
\emph{after} state spheres must avoid intersecting each other
\cite{Schneider-emmgeo2010},
as expressed by
$P > N$
(equation~\eq{eqn.P.N}).
The two constraints on the dimensionality therefore come from
the \emph{after} state spheres
bumping into each other
and from them being compressed within the larger \emph{before} sphere.

We found that when a DNA binding protein evolves to be optimally
efficient, the upper and lower dimensional bounds converge to twice
the information content of the binding site as measured in bits
(\fig{fig.effcurvebounds}).
Using this result, we found that the common $6$ base pair recognizing
restriction enzymes, which require $12$ bits to describe their pattern,
use a $24$ dimensional coding space.
When EcoRI is bound to a DNA sequence
its state can be described as a sphere in
the high dimensional coding space with each
of the possible $4^6=4096$
hexamer sequences represented
by a different sphere~\cite{Schneider.ccmm}.
If the sphere for EcoRI bound to GAATTC were to
overlap with any other sequence sphere, then EcoRI could bind to
and cut at inappropriate locations
that are unprotected by the corresponding methylase, leading to
death~\cite{Heitman.Model1989}.
Since EcoRI binding to sequences other than
GAATTC is at least $10^6$ fold down in
digestion~\cite{Lesser.Jen-Jacobson1990},
these spheres effectively do not intersect.
Excess binding energy that retains the same binding pattern
will be lost by mutational changes in the EcoRI protein structure,
so the spheres must be tightly packed
together in the $24$ dimensional space.
Remarkably,
it has already been shown by coding theorists
that the best known sphere packing is the Leech lattice in
$24$ dimensions~\cite{Conway.Sloane1998, Sloane1998, Cohn.Viazovska2017}.
Likewise the $4$ base pair restriction enzymes
use a $16$ dimensional coding space,
and there are good packings known in that space
(\fig{fig.restpack})~\cite{Conway.Sloane1995}.
Thus, there is
a correlation between commonly observed restriction enzyme
DNA site sizes and the best packing of spheres in high dimensional
spaces.
Could this be a coincidence?
We believe it is not for the following additional reasons.

First, the data sets are large.  The entire collection represents
nearly $4300$ %
restriction enzyme sites
(\tablereference{enzymetable}).
Restriction enzymes are initially discovered by their ability
to digest DNA, and this method does not indicate the sequence of the binding
site, which is unknown until after the enzyme has been
isolated, purified, and characterized.
Odd classes of sites are noticed and publicized because these
are eagerly sought as research reagents.
Likewise,
the data on different kinds of  high dimensional
sphere packings
(\fig{fig.restpack}b)
represent research efforts
spanning the $70$ years since Shannon's publication in 1948,
and there are strong economic
incentives to discover and publicize new
packings because they can be used to improve communications.

Second, the correlation between sphere packing and restriction enzymes
was derived without introducing any free parameters to the boundary equations
for the dimensionality.
It is a natural consequence of previously established molecular machine
theory~\cite{Schneider.ccmm, Schneider.edmm, Schneider-emmgeo2010,
Schneider-brmit2010}.

Presumably natural systems discovered the Leech lattice long ago,
but the details of how a small protein can implement
such a code are unknown.
However,
the fact that restriction enzymes have apparently discovered
good codes
should help us to understand
how they can recognize
short DNA sequences so precisely.
Conversely,
understanding the molecular mechanism of restriction enzyme decoding
could lead to single-molecule communications
devices~\cite{Schneider.ccmm}.

The distribution of restriction enzyme choice of dimensionality
is well explained by the best packings of spheres in various
dimensions
(\fig{fig.restpack}).
The major peak of $6$ base cutting restriction enzymes is
most likely explained
by their use of the Leech lattice in $6 \times 4 = 24$ dimensions.
Likewise, the peaks at $16$ and $32$ dimensions correspond to the $4$
and $8$ base cutters respectively.
In addition, restriction enzyme use of $18$, $20$, and $28$ dimensions
appears to correspond to good nonlattice packings that
are known in those dimensions.
This leaves three holes in the distribution at $22$, $26$, and $30$
dimensions which are rarely used by restriction enzymes but which
have decent sphere packings.
We suggest that restriction enzymes with dimensions close to a peak
evolve into the peak.
For dimension $30$,
\tablereference{enzymetable}
gives the example of KpnBI
with the recognition sequence
CAAANNNNNNRTCA.
Notably this is an asymmetric recognition sequence with
a part that recognizes exactly 4 bases
CAAA
on the $5'$ side and
RTCA on the $3'$ side.
Recognition of a purine R is typically accomplished by
a single hydrogen bond to the N7 position of either A or G
\cite{Seeman.Rich1976, Schneider.oxyr}.
If an additional contact or hydrogen bond into the major groove evolves
(for example to
the N6 of A
or
O6 of G
and on the complementary strand the
methyl of T,
O4 of T
or
N4 of C),
then the enzyme could specify exactly A or G and the dimensionality
would increase to $32$.
This could improve the sphere packing
according to our current knowledge of lattices in $30$ and $32$ dimensions,
so the lack of $D = 30$ enzymes is likely
to be because most have already evolved to the nearby better packing.
Since there are two known $D = 30$ enzymes according to 
\tablereference{enzymetable},
we can test this idea by inspection of the other one.
Indeed, that enzyme is
Eco851I with recognition sequence GTCANNNNNNTGAY.
Since Y pairs to R on the complementary strand,
alteration of the terminal Y to a specific base would switch
this enzyme from $30$ to $32$ dimensions by the same mechanism.
Indeed, a similar explanation for an evolutionarily easy switch
from $26$ to the $24$ dimensional Leech lattice is suggested by the enzyme
RsrII CGGWCCG,
while
the $22$ dimensional
PpuMI RGGWCCY
has three such opportunities.
In each of these cases, merely having the disfavored dimension leads
to a pattern vulnerable to evolution to a nearby dimension.
Whether there are biological or mathematical constraints that prevent
the rare cases from evolving to the best packing dimensions is
unknown.

Two additional biological constraints on the distribution
of restriction enzyme dimensionalities were mentioned earlier.
We expect few if any restriction enzymes to
function at low dimensionality (below $4$ bases or $16$ dimensions) since 
such enzymes would digest DNA frequently and so would
require extensive methylation protection which may be disadvantageous.
Restriction sites longer than $8$ bases or $32$ dimensions would
only be found rarely on invading DNA and so presumably these too
would not have much advantage.
These factors limit the range of functionally useful dimensions.

\section{Coding space as a fitness landscape}
\label{sec:fitness.landscape}

Considering how well the 
restriction enzyme frequency
(\fig{fig.restpack}A)
and
best sphere packing center density
(\fig{fig.restpack}B)
distributions match overall given the biological constraints
on the range that restriction enzymes can function in,
the sphere packing center density
distribution appears to be
a measure of fitness 
for restriction enzymes 
evolving over
a high dimensional
adaptive fitness landscape,
similar to the high dimensional spaces described by
Wright~\cite{Wright1932, Gavrilets1997, Gavrilets2010,
Pigliucci2008}.

For biological systems, the number of sphere centers
corresponds to the number of distinct states the system can be in.
The center density $\delta$
(equation \eq{eqn.littledelta})
is therefore a more appropriate measure than the filled
volume of
the lattice defined by $\Delta$
(equation \eq{eqn.bigDelta})
since biological systems evolve
to have distinct states~\cite{Schneider-emmgeo2010}.
The volume of the state is itself irrelevant.
So we propose that
$\log_2 \delta$ vs.\ $D$ represents the biological coding landscape.
Plotting $\log_2\delta$ instead of $\delta$
emphasizes the detailed features of the curve.
A biological system can evolve to
obtain the highest number of distinct states
by maximizing the sphere center packing density $\delta$.
Because the capacity is the logarithm of the number of states,
this also maximizes the information and the efficiency.

Since the logarithm is monotonic, if
$\delta_1 > \delta_2$, $\log_2{\delta_1} > \log_2{\delta_2}$.
Examining~\fig{fig.restpack}B, we notice several important features.
For lattice packings,
since $\log \delta \le 0$,
there is no more than one center per
unit sphere packing volume for $D\in [0,30]$.
There is exactly one in $D=0$ and $D=24$:
these are the densest sparse packings.
In higher dimensions ($D>30$),
there can be more than one center per
unit sphere packing volume.

Comparing~\fig{fig.restpack}A to~\fig{fig.restpack}B,
we encounter a puzzle.
The center density in $D=20$ is larger than
the center density in $D=16$.
Why then are there more $4$ base cutters than
$5$ base cutters?
There are at least three explanations.
Evolutionary selection
should increase
the molecular machine
capacity by finding relative maxima
of the sphere packing center density
but evolution also
minimizes the expenditure of resources by a cell.
On average, an
enzyme that recognizes a shorter sequence requires less protein
structure
and so requires less energy
to synthesize.
Perhaps
the gain in information density at dimension $20$ compared to dimension $16$
is insufficient to offset the greater energy cost.
In contrast,
the significant
improvement of information density at dimension $24$ may provide a
superior benefit to the organism despite the extra expenditure in
energy and this may explain why there are more $6$ base enzymes than $4$ base
enzymes.
A second consideration
is the number of target restriction sites needed
on foreign DNA.
Larger target DNAs would be best digested less frequently
so that the restriction enzyme spends less time on small regions
and conversely some enzymes may be targeted to smaller DNAs,
leading to a smaller dimensionality.
A third factor is the ease of evolving recognition patterns.
Protein
dimerization allows the creation of a 4 base cutting restriction site
from two half sites.
Five base recognition is probably more difficult
to evolve
since the central base has to be handled separately or the entire
site has to become asymmetric.

The best center density $\delta$ in $16$ dimensions has the
same value as the best center density in 
$17$ dimensions
(\fig{fig.restpack})~\cite{Conway.Sloane1998}.
Since higher dimensionality allows lower error rates~\cite{Shannon1949},
why isn't $17$ dimensions used
to provide more accurate restriction?
A $17$ dimensional packing would take $4.25$ bases and the $0.25$
bases could be contacted in the center of the site.
Dimeric proteins use less protein structure than a monomer and so
require smaller DNA coding,
but
only one of the two monomers
could contact the center at a time.
Perhaps this awkward wasteful situation is unfavorable compared
to using $16$ dimensions.
In fact, odd dimensions are avoided by restriction enzymes in general
(\tablereference{enzymetable}).
Another possibility is that the huge number of known packings
in $16$ dimensional space
($10^7$,
\cite{Conway.Sloane1995})
overwhelms a smaller number of packings in $17$ dimensions.
Similar considerations apply to dimensions $7$ and $8$,
where the center densities are equal.
Here, because $8$ is even,
the $E_8$ lattice is known to be highly symmetric,
and the error rates are smaller, the preference for
$D=8$ over $D=7$ is clear.

According to \fig{fig.restpack}A there are several hundred restriction
enzymes that operate in $16$, $18$ and $20$ dimensional coding spaces.
Just as the $E_8$ and Leech lattices allow for dense sphere packings
in $8$ and $24$ dimensions, evolution may select similarly dense
packings in other even dimensions, especially those divisible by $4$. 
There may be best packings in these dimensions that have not been
discovered yet; this is an open problem in mathematics.

Evidently many restriction enzymes have discovered the Leech lattice,
but does this merely reflect divergent evolution from a common
ancestor?
Many restriction enzymes
have widely different
structures~\cite{Pingoud.Wende2014}, %
suggesting convergent evolution.
Perhaps a deeper understanding of the coding spaces
will help to classify these enzymes.
In addition,
the coding spaces of restriction enzymes may provide
a fertile ground for precise quantitative analysis of population genetics
and theoretical evolutionary biology
since much is known about sphere packing in high
dimensions~\cite{Conway.Sloane1998}.
Though
we have found a strong
correlation between the high dimensional sphere center packing landscape
and restriction enzyme information content preferences,
this still leaves the important task of understanding
how the codes are implemented by the protein structures
as a major problem for coding theorists and biologists.

\section{Transcription factors use high dimensional fractal coding}
\label{sec:transcription.factors}

Experimental evidence has been obtained indicating that
for the transcription factor Fis,
the specific DNA binding mode differs sharply from
non-specific DNA binding
since there is a break in the 
binding curve at zero bit binding
sites~\cite{Schneider.edmm, Schneider-Ri1997, Shultzaberger.Schneider-spr2007}.
Shannon pointed out that a mapping from a high dimensional
space to lower dimensions creates discontinuities~\cite{Shannon1949},
so this
result suggests that Fis functions in a high dimensional
coding space
close to $2 \times 7.86 \pm 0.27 = 15.72 \pm 0.54 \approx 16$
dimensions~\cite{Hengen.Schneider-fisinfo1997}.

For nucleic acid recognizing molecules
that have specific sites on the genome,
such as transcription factors,
the information
in the binding sites, $\rsequence$, evolves to match
the information needed to locate the binding sites,
\begin{equation}
\rfrequency = - \log_2 \gamma / G
\;\;\;\;\;\mbox{(bits per molecular operation)}
\label{eqn.rf}
\end{equation}
where $G$ is the number of potential
binding sites in the genome
and $\gamma$ is the number of specific binding
sites~\cite{Schneider.Ehrenfeucht1986, Schneider.ev2000}.
In the case of restriction enzymes equation~\eq{eqn.rf} does
not apply since there are no specific binding sites in the foreign
DNA attacked by these enzymes
and $\gamma$ does not have a particular value.
However, the principle that
$\rsequence \approx \rfrequency$
does apply to many other genetic systems
such as
transcription factors~\cite{Schneider.Ehrenfeucht1986},
promoters~\cite{Shultzaberger.Schneider-flexprom2007, Penotti1990},
ribosome binding sites \cite{Shultzaberger.Schneider2001, Schneider2005},
and mRNA splicing~\cite{Stephens.Schneider-splice1992}.
In general
$G / \gamma$ is not an exact power of two,
so $\rfrequency$, and therefore $\rsequence$ is usually not an integer.
Equation~\eq{eqn.lower.upper.D.bound}
then implies that
for an optimal molecular machine in which $D = 2R$,
the dimension of the binding sites
will not be an integer.
Objects with non-integer dimensions are called
fractals~\cite{Mandelbrot1983, Muslih.Agrawal2009, Sorensen.Roberts1997}.
As shown in
\tablereference{enzymetable}, many restriction enzymes
may also have fractal dimensions, although more close inspection
using sequencing technologies may be required to confirm the
observation~\cite{Cohen-Karni.Zheng2011}.
How a molecular coding system can have non-integer dimensions and
the possible applications of
such high dimensional fractal codes for communication systems
remain to be investigated.

\section*{Acknowledgments}

TDS thanks
N.~J.~A.~Sloane for useful discussions about the normalization
of the sphere packing density function,
Rich Roberts
for useful discussions about REBASE
and
for pointing out that Type I restriction enzymes differ from Type II,
Michael Smith
for suggesting that sphere packing density may be an adaptive landscape,
Eckart Bindewald,
Misha Kashlev,
Ryan Shultzaberger
and
Randall Johnson
for comments on the manuscript
and the Advanced Biomedical Computing Center (ABCC) for support.
VJ thanks the U.S.~National Cancer Institute
Werner H. Kirsten Student Intern Program,
Soren Brunak, and the
Technical University of Denmark for hospitality during initial stages
of this project in 1994.
\textbf{Data and materials availability:}
see
Table \ref{enzymetable}
and
\fig{fig.restpack}.

\nolinenumbers

\raggedright
\bibliography{alldoi}

\end{document}